\documentclass[11pt,twoside]{article}
\usepackage{./asp2014}

\aspSuppressVolSlug
\resetcounters

\begin{document}

\title{Science with an ngVLA: The Sunyaev-Zeldovich Effect from Quasar and Starburst Winds.}
\author{Mark Lacy$^1$, Suchetana Chatterjee$^2$, Avinanda Chakraborty$^2$, Brian Mason$^1$, Craig Sarazin$^3$, Amy Kimball$^4$, Kristina Nyland$^1$, Graca Rocha$^5$ and Barney Rowe$^6$
\affil{$^1$NRAO, Charlottesville, VA, USA; \email{mlacy,bmason,knyland@nrao.edu}}
\affil{$^2$Presidency University, Kolkata, India}
\affil{$^3$Department of Astronomy, University of Virginia, 530 McCormick Rd., Charlottesville, VA 22904, USA}
\affil{$^4$NRAO, Socorro, NM, USA}
\affil{$^5$Jet Propulsion Laboratory, Pasadena, CA, USA}
\affil{$^6$University College, London, UK}}

% This section is for ADS Processing.  There must be one line per author.
\paperauthor{Sample~Author1}{Author1Email@email.edu}{ORCID_Or_Blank}{Author1 Institution}{Author1 Department}{City}{State/Province}{Postal Code}{Country}
\paperauthor{Sample~Author2}{Author2Email@email.edu}{ORCID_Or_Blank}{Author2 Institution}{Author2 Department}{City}{State/Province}{Postal Code}{Country}
\paperauthor{Sample~Author3}{Author3Email@email.edu}{ORCID_Or_Blank}{Author3 Institution}{Author3 Department}{City}{State/Province}{Postal Code}{Country}

%Chapters do not have abstracts
\begin{abstract}
The Sunyaev-Zeldovich Effect (SZE) can be used to detect the hot bubbles in the intergalactic medium blown by energetic winds from AGN and starbursts. By directly constraining the kinetic luminosity and total energy of the outflow, it offers the promise of greatly increasing our understanding of the effects of wind feedback on galaxy evolution. Detecting the SZE in these winds is very challenging, at the edge of what is possible using existing facilities. The scale of the signal (10-100~kpc) is, however, well matched to interferometers operating at mm wavelengths for objects at $z\sim 1$. Thus this could become a major science area for the ngVLA, especially if the design of the core is optimized for sensitivity on angular scales of $>1$ arcsec in the 90~GHz band.
\end{abstract}

\section{Introduction} \label{sec:intro}

Winds from AGN and starbursts are one of the major sources of feedback in galaxy evolution. Their interaction 
with the interstellar medium (ISM) of the galaxy can inject turbulence, dissociate molecular gas, or even drive the gas out of the galaxy completely (Silk \& Rees 1998; Bower et al.\ 2006; Hopkins et al.\ 2006; Croton et al.\ 2006; Richardson et al.\ 2016). Estimating the energy in a wind or outflow is very difficult, however, as, in most models, the predominant phase in the outflowing gas is hot ($\sim 10^7$K) gas with low density. Emission from such gas can only be detected in the X-ray. The dependence of X-ray emission on the square of the density, and confusion with point sources such as AGN, make any detection extremely difficult (e.g., Powell et al. 2018). There is another way to detect the host gas in these winds other than through their emission though, and that is through the Sunyaev-Zeldovich Effect (SZE). This is the distortion to the spectrum of the cosmic microwave background (CMB) radiation when it travels through hot gas. Two terms exist, one thermal and one kinetic. The thermal component (the tSZE) is proportional to the line of sight electron pressure through the gas, while the kinetic component (the kSZE) is proportional to the line of sight bulk velocity of the electrons. While the tSZE is now routinely detected in galaxy clusters (e.g., Bleem et al.\ 2015), the kSZE has only been convincingly detected in the cluster MACS J0717.5+3745 (Sayers et al. 2013; Adam et al. 2017). 

A wind from a quasar or starburst produces an expanding bubble of hot gas that is highly overpressured compared to the surrounding medium (interstellar, intragroup or intracluster), and moving with respect to the frame of the microwave background (Figure 1). 
Thus, as first suggested by Natarajan \& Sigurdsson (1999), it should be possible to 
detect the SZE towards these winds (Yamada \& Fujita 2001; Chatterjee \& Kosowsky 2007; Chatterjee et al.\ 2008; Rowe \& Silk 2011; hereafter RS11). Statistical studies using stacked data from single dish telescopes have detected significant signals from quasar hosts (Chatterjee et al. 2010; Ruan et al. 2015; Crichton et al.\ 2016), but it is unclear whether these results are affected by contamination of the SZE signal within the large single beam size by either the intragroup or intracluster medium around the quasar (below the null in the tSZE at 218~GHz; Figure 2), or from star formation in the quasar host at high frequencies (above 218~GHz) (Cen \& Safarzadeh 2015; Soergel et al.\ 2017). The SZE signal is independent of redshift, making it a powerful tool for studying the intergalactic medium (IGM) at high redshifts.

\begin{figure}
\label{fig:szschematic}
\centering
\includegraphics[width=4.0in]{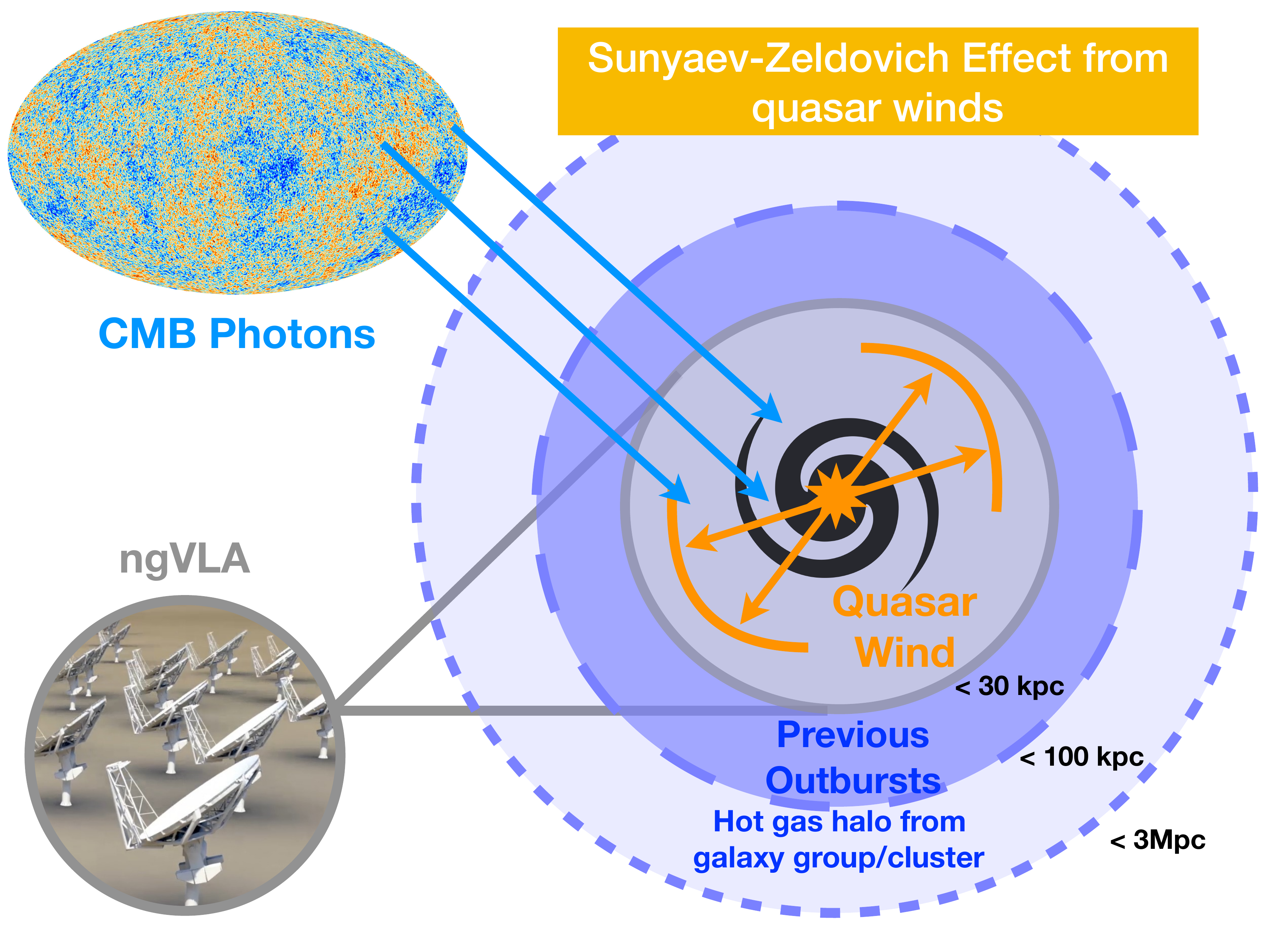}
\caption{Schematic of the observation of a quasar wind using the ngVLA. Microwave background photons are upscattered in energy through their interaction with fast moving electrons in an expanding quasar wind. Contributions come both from the bulk motion of the electrons towards or away from the observer, and from the thermal motion of the electrons in the hot wind. Relics of previous activity also contribute to the SZE on scales $\sim 100\,$kpc, along with any hot gaseous halo from a group or cluster of galaxies surrounding the quasar on Mpc scales. }
\end{figure}

\begin{figure}
\label{fig:szspec}
\centering
\includegraphics[width=4.0in]{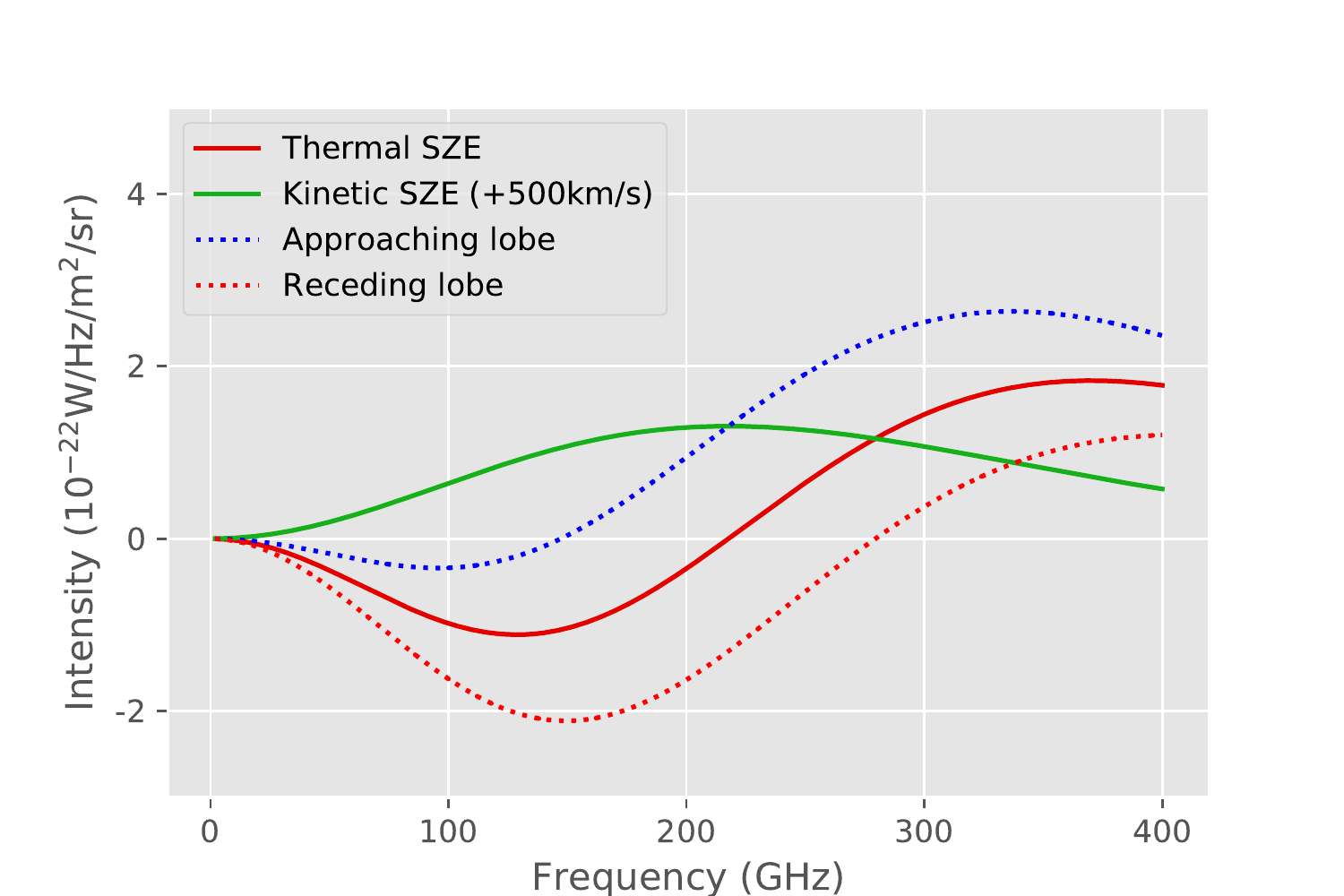}
\caption{The intensity of the SZE as a function of frequency. The purely thermal model (red) assumes 
a Compton Parameter $y=10^{-5}$ (see Section \ref{sec:calcs}), and shows the typical tSZE signal from a quasar wind bubble, with a decrement in the microwave background observed shortward of 218~GHz and an increment at higher frequencies. The kinetic model (green) is the signal from electrons with a bulk velocity of 500~km/s towards 
the observer with the same line of sight optical depth as the thermal population (assuming a temperature of $10^7$K
for the thermal elections). The two dashed lines show the situation when the outflow consists of two lobes, one approaching the observer and one receding (Lacy et al.\ 2018). The thermal decrement signal from the receding lobe is enhanced by the bulk motion of the outflow. For the approaching lobe, the decrement signal is almost canceled by the kinetic signal, but the increment at high frequencies is enhanced.}

\end{figure}

\section{Direct detection experiments}

Attempts to directly detect the SZE from quasar- or starburst-driven winds can be made with interferometers such as the Atacama Large Millimetre/submillimetre Array (ALMA) (e.g. Chatterjee et al.\ 2008; RS11), though the observations are very challenging, requiring long (tens of hours) integration times to detect or constrain even the most powerful outflows. At $z\stackrel{>}{_{\sim}}1$, the predicted size scale of the SZE signal, $\sim 10-100\,$kpc, is well matched to the angular resolution of ALMA or the VLA in their most compact configurations, and thus to the core of the proposed ngVLA. The good match of the interferometer beam to the expected scale of the signal reduces the effect of beam dilution compared to single dish observations (though the largest outflows will be well matched to the resolution of the Greenbank Telescope (GBT), $\approx 10^{''}$ at 90~GHz). The need to subtract contaminating emission from the central object (AGN or starburst) driving the flow also means that high angular resolution is needed. This can be achieved either by taking data in a larger configuration, or, in the case of a relatively well-filled aperture like ALMA or the ngVLA core, weighting the visibilities in such a way as to enhance the resolution. 

\subsection{The contribution of the Kinetic SZE}

In galaxy clusters, the kSZE, due to the motion of the cluster relative to the CMB, is typically much smaller in magnitude than 
the tSZE from the thermal motions of the electrons, and is hard to disentangle from the tSZE signal. In winds though, the kSZE signal can be very strong as the bulk motion of the electrons can be several thousand ${\rm km\,s^{-1}}$. Even in post-shocked gas, the bubble expansion rate can be $>>100\,{\rm km\,s^{-1}}$ and contribute significantly to the SZE signal (see discussion in Lacy et al.\ 2018). Furthermore, because the sign of the kSZE depends on the bulk velocity of the electrons relative to the line of sight, in a bipolar outflow one side will have the tSZE and kSZE adding together, and one side will have them subtracting, resulting in asymmetric 
decrements at the observing frequencies of the ngVLA (Figure 2). To provide further evidence of a kSZE detection, complementary observations above the tSZE null at 218~GHz will be needed to see the expected change in sign of the tSZE signal. Detection of both the kSZE and tSZE signal will allow the accurate characterization of the hot wind, including its kinetic luminosity, age, and mass outflow rate.

\subsection{Relativistic jets}

Feedback from relativistic plasma jets (as opposed to thermal winds), in the ``radio'' mode (e.g., Fabian 2012) has also been suggested as an important influence on galaxy formation. The potentially long lifetime of the low-power radio jets powering radio galaxies with luminosities $\sim 10^{24}$WHz$^{-1}$ ($\stackrel{>}{_{\sim}} 10^8$ yr)  makes this form of feedback an attractive option for heating the intragroup or intracluster medium consistently during gaps between bursts of AGN activity at high accretion rates ($\sim 0.1-1\times$Eddington). Feedback by short-lived ($\sim 10^7$\;yr) powerful FRII radio jets (e.g., Nesvadba et al.\ 2017) is easy to detect (though 
understanding its effects on the ISM of the host is more complex, e.g., Mukherjee et al.\ 2016), but the radio mode
typically involves radio sources of much lower power ($L_{\rm 1.4GHz}\sim 10^{24}{\rm WHz^{-1}}$). %Furthermore, 
%if the AGN is not currently producing jets, the radio emission from ``relic'' radio sources fades 
%very quickly, in $\sim 10^{7-8}$yr, and such sources can often be seen only by the bubbles they excavate in a %surrounding intracluster medium (e.g., Birzan et al.\ 2008). 

As electrons become relativistic, the strength of the SZE as a function of electron pressure drops. So a relic radio source can be detected as a local reduction of the SZE signal in the intracluster medium. This opens up the possibility of finding evidence for ancient ($>10^8$yr) episodes of radio mode feedback via the SZE (Pfrommer et al.\ 2005), even if the synchrotron emission has faded. Indeed, as the morphology and scale size of any thermal wind signal is likely to differ from that of a well-collimated relativistic jet, it might be possible to detect both an enhancement in the SZE from a thermal wind and a reduction from bubbles fed by relativistic plasma from radio jets in a single object. This makes the lower frequency capabilities of the ngVLA important to this case too - sensitive observations at cm wavelengths and subarcsecond resolution will be needed to image any jets so we can understand the relative contributions of thermal winds and relativistic jets to the feedback process.

\begin{figure}
\includegraphics{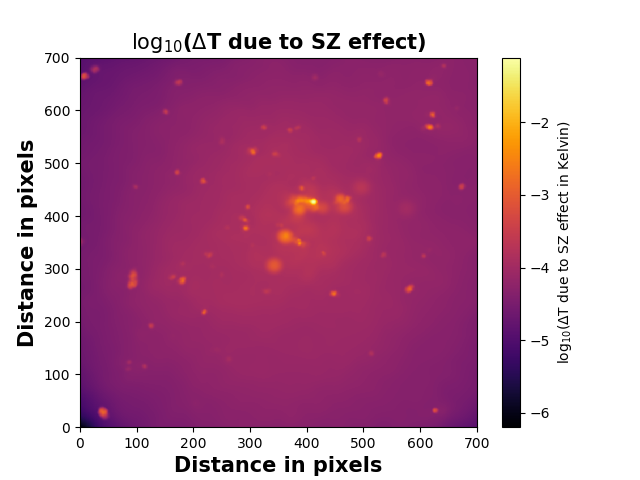}
\caption{Simulation of the tSZE signal from Chakraborty (2017) using the hydrodynamic simulations of Khandai et al. (2015). The central object is a $2\times 10^9 M_{\odot}$ quasar with a bolometric luminosity of $3\times 10^{11} L_{\odot}$. Each pixel is 0.14~kpc in size; the image is $\approx $100~kpc on a side.}
\end{figure}

\section{Experimental Design}

\subsection{Signal Strength and Scale Size}\label{sec:calcs}

The intensity of the SZE spectral distortion observed at frequency $\nu$ (wavelength $\lambda$), assuming an isothermal plasma and defining $x = h \nu/k T_{\rm CMB}$, is 
given by e.g.\ Sazonov \& Sunyaev (1998):
\begin{equation}
\Delta I(x) = \frac{2 k T_{\rm CMB}}{\lambda^2} \frac{x^2 e^x}{(e^x-1)^2} \tau \left( \frac{k T_e}{ m_e c^2} f_1(x) +  \frac{v_r}{c}  \right).
\label{eq:sz}
\end{equation}
The left-hand term inside the brackets on the right-hand side of Equation \ref{eq:sz} represents the tSZE component, and the right-hand term the kSZE.
$\tau$ is the Thompson scattering optical depth through the plasma, $\tau = \sigma_T \int dl \, n_e(l)$;  and $f_1(x) = x \, {\rm coth}(x/2)-4$  expresses the frequency dependence of the tSZE. 
%In the low-frequency limit the thermal SZE can be shown to give $\Delta I/I = \Delta T/T = -2 y$.
The $v_r/c$ term due to the kSZE is proportional to the line of sight velocity $v_r$. This expression neglects higher order terms in $v_r/c$ and $k T_e / m_e c^2$. The leading order relativistic correction in $T_e$ is a $4\%$ correction to the tSZE magnitude for a $T_e =  \ 5 \, {\rm keV}$ plasma at $\nu = 140 \, {\rm GHz}$ (Itoh \& Nozawa 2004). The strength of the tSZE is often expressed in terms of the Compton parameter,
$y=\tau k_{\rm B} T_{\rm e}/(m_{\rm e}c^2)$. 

Both analytical and numerical models (e.g., Figure 3) provide some guidance on the likely size scale and detectability of the tSZE from winds. RS11 give an analytic prescription for starburst winds using a simple adiabatic model (which works equally well for AGN winds as the model assumes a spherically symmetric outflow from a small central source). They obtain a peak value for the $y$-parameter in the bubble, 
\begin{equation}\label{eqn:yparam}
y_{\rm peak}=1.13\times 10^{-5} (L_{\rm W}/10^{12}{\rm L_{\odot}})^{3/5}(t/10^7{\rm yr})^{1/5}(1+z)^{6/5},
\end{equation}
where $L_{\rm W}$ is the kinetic power of the wind, and $t$ the age of the outflow, with an estimated radius of the bubble of
\begin{equation}\label{eqn:size}
R=48 \left( \frac{L_{\rm W}}{10^{12} {\rm L_{\odot}}} \right)^{1/5} \left(\frac{t}{10^7 {\rm yr}}\right)^{3/5} (1+z)^{-3/5} {\rm kpc}.
\end{equation}

The gas temperature in the post-shock region behind the edge of the bubble, $T_s = 3 \mu m_p v_s^2 / (16 k)$, is typically high ($\sim 10^7$K), and the densities low ($\sim 10^{-5} {\rm m^{-3}}$), resulting in long cooling times $\sim 10^9$\,yr. The contribution from the kSZE is dependent on geometry and
bubble expansion speed, as well as the observing frequency, but can be comparable to that from the tSZE (Lacy et al.\ 2018). 

Hydrodynamic simulations by Chatterjee et al.\ (2008) emphasize the cumulative effect of outbursts over the lifetime of the host galaxy. These indicate a larger halo of hot gas, $\sim 0.1-1$~Mpc in size and formed from the superposition of multiple AGN outbursts, that can be detected by telescopes with high surface brightness sensitivity and less filtering of emission on large angular scales than the ngVLA interferometer. For this science case in particular, a large single dish addition to the ngVLA would be optimal.

The peak signal from the tSZE decrement (which is relevant for the ngVLA
frequency range) is at $\approx130$~GHz, higher than the proposed maximum operating frequency (though accessible to ALMA). The very large collecting area of the ngVLA, however, will make it more effective than ALMA at this science even operating at 100~GHz, where the tSZE decrement signal remains 87\% of the peak.

\subsection{Observing Strategy and possible targets for the ngVLA}

As the strength of the tSZE signal is proportional to the wind luminosity to the 3/5 power, it makes sense to target the most luminous examples of starbursts and quasars known. 
%Equation \ref{eqn:yparam} also suggests high redshift should enhance the signal; however, this applies only to a single event, and at lower redshifts we should see multiple events superposed, enhancing the signal. 
The size scale of the wind bubble of a few arcsec makes it well suited to interferometers rather than the single dishes used in most SZE experiments, and is only a weak function of the luminosity of the wind, mostly depending on the age of the outflow.

Possible starburst targets are discussed in RS11. Quasar targets include the most luminous quasars in the northern hemisphere, with bolometric luminosities ($L_{\rm bol}$) close to $10^{15}L_{\odot}$ (e.g.\ HS~1700+6421; $z=2.75$). Another strategy would be to target luminous quasars with known winds in their absorption line spectra (Broad Absorption Line quasars). Even using these luminous targets, detection of any signal will be challenging. For the hyperluminous southern quasar HE~0515-4414 (which also has $L_{\rm bol}\approx 10^{15}L_{\odot}$), we detected a signal at 3.5$\sigma$ with a $y_{\rm peak}=2\times 10^{-5}$ using a deep ALMA (12.4~hr time on source) observation. This detection corresponds to a kinetic luminosity in the winds of about 0.01\% of $L_{\rm bol}$ (Lacy et al.\ 2018). Nevertheless, the outflow in HE~0515-4414 still has a total enthalpy $\approx 3\times 10^{53}$W, comparable to the largest bubbles excavated by radio jets in radio mode feedback, and also seems to be long-lived (with a cooling time $\sim 0.6$~Gyr), and so could be providing feedback in a similar manner to radio sources. 
%Thus detections/limits below $y\sim 10^{-5}$, only realistically achievable at arcsecond resolution with the ngVLA, are needed to be able to assess the role of these bubbles in feedback.

For a given $\tau$, $T_e$ and $v_{r}$, the SZE signal from winds is independent of redshift, making these observations particularly valuable for studying winds at $z>>1$, which would be almost impossible to detect in the X-ray. The simple RS11 wind model in Equations \ref{eqn:yparam} and \ref{eqn:size} suggest that a wind at high redshift would have both a higher decrement and a smaller size (making it more suitable for follow-up with an interferometer). However, the RS11 model assumes that the IGM density around the quasar is scaling only with redshift, which may not be the case in practice, and also, at higher redshift, the signal is less likely to have been enhanced by earlier outbursts of activity. Further observations are thus needed to investigate how the detectability of the SZE signal from winds depends on redshift.

Yet another approach is to select the most massive quiescent elliptical galaxies that must have supported luminous starburst and/or quasar activity in the past on the basis that -- assuming a fixed fraction of the radiative energy goes into the wind -- then these object will have the largest remnant bubbles, even if they are not currently active. An example of this approach applied to a stacking analysis is the 3.6$\sigma$ detection of an SZE signal in stacked luminous red galaxies (Spacek et al. 2016). 

The larger collecting area and wide bandwidth of ngVLA make it potentially much more sensitive than ALMA for detecting these winds, however, the current proposed configuration (``ngvla-revB''), where the core is offset to the north of the overall antenna distribution is not optimal for sensitivity on angular scales $>1$-arcsec at 100\,GHz (Carilli 2018), so in practice the detection sensitivity to a 50~kpc outflow signal at $z\sim 1$ is comparable to ALMA. Where the ngVLA will come into its own is in spatially resolving the most powerful outflows, where $\approx 2$ arcsec resolution will be possible at 41~GHz without severe sensitivity losses from tapering. This will allow the accurate measurement of wind properties in the most energetic outflows, and a detailed comparison of the feedback from thermal winds versus non-thermal radio jets.

\acknowledgements SC and AC would like to thank Rudrani Kar Chowdhury and Alankar Dutta for help with the simulation code. The National Radio Astronomy Observatory is a facility of the National Science Foundation operated under cooperative agreement by Associated Universities, Inc.
% Keep this text on the same line as the \verb"\acknowledgements" command because it makes things a lot easier.

%\bibliography{editor}  % For BibTex

% For non-BibTex:

\end{document}